\documentclass[review,5p]{elsarticle}
\usepackage{graphicx} 
\usepackage{hyperref} 
\usepackage{subfigure}
\usepackage{epstopdf}

\begin{document}


\title{Some New Results on Charged Compact Boson Stars}

\author[du]{Sanjeev Kumar}
\ead{sanjeev.kumar.ka@gmail.com}
\author[duk,iowa,ifp]{Usha Kulshreshtha}
\ead{ushakulsh@gmail.com, ushakuls@iastate.edu}
\author[du,iowa,ifp]{Daya Shankar Kulshreshtha}
\ead{dskulsh@gmail.com, dayakuls@iastate.edu}
\author[ifp]{Sarah Kahlen}
\ead{sarah.kahlen@uni-oldenburg.de}
\author[ifp]{Jutta Kunz}
\ead{jutta.kunz@uni-oldenburg.de}
\address[du]{Department of Physics and Astrophysics, University of Delhi, Delhi-110007, India}
\address[duk]{Department of Physics, Kirori Mal college, University of Delhi, Delhi-110007, India}
\address[iowa]{Department of Physics and Astronomy, Iowa State University, Ames, 50010 IA, USA}
\address[ifp]{Institut f\"ur Physik, Universit\"at Oldenburg, Postfach 2503, D-26111 Oldenburg, Germany}

\date{\today}

\begin{abstract}
In this work we present some new results obtained 
in a study of the phase diagram of charged compact boson stars 
in a theory involving a complex scalar field with a conical potential
coupled to a U(1) gauge field and gravity. 
We here obtain new bifurcation points in this model. 
We present a detailed discussion of the various regions 
of the phase diagram with respect to the bifurcation points. 
The theory is seen to contain rich physics in a particular domain 
of the phase diagram.
\end{abstract}

\maketitle

In this work we study the phase diagram of charged compact boson stars 
in a theory involving a complex scalar field with a conical potential
coupled to a U(1) gauge field and gravity
\cite{Kleihaus:2009kr,Kleihaus:2010ep}. 
A study of the phase diagram of the theory yields new bifurcation points 
(in addition to the first one obtained earlier, 
cf.~Refs.~\cite{Kleihaus:2009kr,Kleihaus:2010ep}), 
which implies rich physics in the phase diagram of the theory. 
In particular, we present a detailed discussion of the various regions 
in the phase diagram with respect to the bifurcation points.

Let us recall that the boson stars 
(introduced long ago \cite{Feinblum:1968,Kaup:1968zz,Ruffini:1969qy}) 
represent localized self-gravitating solutions 
studied widely in the literature 
\cite{Jetzer:1991jr,Lee:1991ax,Mielke:2000mh,Liebling:2012fv,Friedberg:1976me,Kleihaus:2009kr,Kleihaus:2010ep,Hartmann:2012da,Hartmann:2012wa,Hartmann:2013kna,Kumar:2016sxx,Kumar:2016oop,Kumar:2014kna,Kumar:2015sia}.

In Refs.~\cite{Kumar:2016oop,Kumar:2016sxx}, 
three of us have undertaken studies of boson stars and boson shells 
in a theory involving a massive complex scalar field 
coupled to a $U(1)$ gauge field $A_{\mu}$ and gravity 
in the presence of a cosmological constant $\Lambda$. 
Our present studies extend the work of 
Refs.~\cite{Kleihaus:2009kr,Kleihaus:2010ep}),
performed in a theory without a cosmological constant $\Lambda$ 
for a complex scalar field with only a conical potential,
i.e., the scalar field is considered to be massless.
Such a choice is possible for boson stars in a theory with 
a conical potential, since this potential yields compact boson star solutions 
with sharp boundaries, where the scalar field vanishes.
This is in contrast to the case of non-compact boson stars,
where the the mass of the scalar field is a basic ingredient
for the asymptotic exponential fall-off of the solutions.

We construct the boson star solutions of this theory numerically. 
Our numerical method is based on the Newton-Raphson scheme 
with an adaptive stepsize Runge-Kutta method of order 4.
We have calibrated our numerical techniques 
by reproducing the work of Refs.~\cite{Kleihaus:2009kr,Kleihaus:2010ep} 
and \cite{Kumar:2014kna,Kumar:2015sia,Kumar:2016oop,Kumar:2016sxx}.

We consider the theory defined by the following action 
(with $V(|\Phi|):=\lambda |\Phi|$, where $\lambda$ is a constant parameter):
\begin{eqnarray}
S&=&\int \left[ \frac{R}{16\pi G}   +\mathcal L_M \right] \sqrt{-g}\ d^4\,x\; ,\label{3:action}
\nonumber\\ \mathcal L_M& =&	- \frac{1}{4} F^{\mu\nu} F_{\mu\nu}
   -  \left( D_\mu \Phi \right)^* \left( D^\mu \Phi \right)
 - V(|\Phi|)\,,\nonumber\\
 D_\mu \Phi &=& (\partial_\mu \Phi + i e A_\mu \Phi)\ \ ,\nonumber\\\ \ F_{\mu\nu} &=& (\partial_\mu A_\nu - \partial_\nu A_\mu) .
\end{eqnarray}
Here $R$ is the Ricci curvature scalar, 
$G$ is Newton's gravitational constant. 
Also, $g = det(g_{\mu\nu})$, 
where $g_{\mu\nu}$ is the metric tensor, 
and the asterisk in the above equation denotes complex conjugation. 
Using the variational principle, the equations of motion are obtained as:
\begin{eqnarray}
G_{\mu\nu}&\equiv& R_{\mu\nu}-\frac{1}{2}g_{\mu\nu}R = 8\pi G T_{\mu\nu}\,,
\nonumber \\ 
 \partial_\mu \left ( \sqrt{-g} F^{\mu\nu} \right)&=&
   -i e \sqrt{-g}[\Phi^* (D^\nu \Phi)-\Phi (D^\nu \Phi)^* ]\,,
\nonumber 
\end{eqnarray}
\begin{eqnarray}
D_\mu\left(\sqrt{-g}  D^\mu \Phi \right)& =& \frac{\lambda }{2}\sqrt{-g}\,\frac{\Phi}{|\Phi|}\,,\nonumber \\
\left[D_\mu\left(\sqrt{-g}  D^\mu \Phi \right)\right]^*& =& \frac{\lambda }{2}\sqrt{-g}\,\frac{\Phi^*}{|\Phi|} .
  \label{3:vfeqH}
 \end{eqnarray}
The energy-momentum tensor $T_{\mu\nu}$ is given by
\begin{eqnarray}
T_{\mu\nu} &=& \biggl[ ( F_{\mu\alpha} F_{\nu\beta}\ g^{\alpha\beta} -\frac{1}{4} g_{\mu\nu} F_{\alpha\beta} F^{\alpha\beta})
\nonumber\\ & & + (D_\mu \Phi)^* (D_\nu \Phi)+ (D_\mu \Phi) (D_\nu \Phi)^*  
 \\ & &  -g_{\mu\nu} \left((D_\alpha \Phi)^* (D_\beta \Phi)    \right) g^{\alpha\beta}
 -  g_{\mu\nu} \lambda( |\Phi|) \biggr] . \ \ \  \nonumber \label{3:vtmunu}
\end{eqnarray}

To construct spherically symmetric solutions 
we adopt the static spherically symmetric metric 
with Schwarzschild-like coordinates
\begin{equation}
ds^2= \biggl[ -A^2 N dt^2 + N^{-1} dr^2 +r^2(d\theta^2 + \sin^2 \theta d\phi^2) \biggr].
\end{equation}
This leads to the components of Einstein tensor ($G_{\mu\nu}$) 
\begin{eqnarray}
G_t^t &=& \biggl[ \frac{-\left[r\left(1-N\right)\right]'}{r^2} \biggr] ,\nonumber \\  
G_r^r &=& \biggl[ \frac{2 r A' N -A\left[r\left(1-N\right)\right]'}{A\ r^2} \biggr] ,\nonumber \\
G_\theta^\theta &=& \biggl[ \frac{2r\left[rA'\ N\right]' + \left[A\ r^2 N'\right]'}{2 A\ r^2} \biggr]
\  \   = \  G_\varphi^\varphi .
\end{eqnarray}
Here the arguments of the functions $A(r)$ and $N(r)$ have been suppressed. 
For solutions with a vanishing magnetic field, 
the Ans\"atze for the matter fields have the form:
\begin{equation}
 \Phi(x^\mu)=\phi(r) e^{i\omega t}\ \ ,\ \ A_\mu(x^\mu) dx^\mu = A_t(r) dt .
\end{equation}

We introduce new constant parameters:
\begin{equation}
 \beta=\frac{\lambda\,e}{\sqrt{2}}   \ \ \ ,\ \ \ \alpha^2 (:=a)= \frac{4\pi G\,\beta^{2/3}}{e^2} . \label{parameters}
\end{equation}
Here $a:=\alpha^2$ is dimensionless.
We then redefine $\phi(r)$ and $A_t(r)$:
\begin{equation}
 h(r)=\frac{(\sqrt{2} \;e\, \phi(r))}{\beta^{1/3}} \ \ \ ,\ \ \  b(r)=\frac{(\omega+e A_t(r))}{\beta^{1/3}} . \label{hb}
 \end{equation}
Introducing a dimensionless coordinate $\hat{r}$ 
defined by $\hat{r}:=\beta^{1/3}\,{r}$
(implying $\frac{d}{d{r}}=\beta^{1/3}\frac{d}{d\hat{r}}$),  
Eq.~(\ref{hb}) reads:
\begin{equation}
 h(\hat{r})=\frac{(\sqrt{2} \;e\, \phi(\hat{r}))}{\beta^{1/3}} \ \ \ ,\ \ \  b(\hat{r})=\frac{(\omega+e A_t(\hat{r}))}{\beta^{1/3}} . \label{hb1} 
 \end{equation}

The equations of motion in terms of $h(\hat{r})$ and $b(\hat{r})$ 
(where the primes denote differentiation with respect to $\hat{r}$,
and ${\rm sign }(h)$ denotes the usual signature function) read:
\begin{eqnarray}
\left[A N \hat{r}^2 h'\right]' &  = & \frac{\hat{r}^2}{AN}\left(A^2N {\rm sign}(h) -b^2 h\right)\ ,  \label{3:vheq}\\
\left[\frac{\hat{r}^2 b'}{A}\right]' &  = & \frac{b h^2 \hat{r}^2}{AN}  . \  \label{3:vbeq}
\end{eqnarray}
We thus obtain the set of equations:
\begin{eqnarray}
N' & = & \bigg[\frac{1-N}{\hat{r}} -\frac{\alpha^2 \hat{r}}{A^2 N}\left(A^2 N^2 h'^2 + N b'^2 \right.\nonumber\\ & &\hspace{1in}\left.+2 A^2 N h+ b^2 h^2\right)\bigg] \ , \label{eq_N}\\
A' & = & \bigg[\frac{\alpha^2 \hat{r}}{A N^2}\left(A^2 N^2 h'^2 + b^2 h^2\right)\bigg] ,\\ \label{eq_A}
h'' & = &\bigg[ \frac{\alpha^2}{A^2N} \hat{r} h' \left(2A^2  h +b'^2\right)-\frac{h'\left(N+1\right)}{\hat{r} N}\nonumber\\&&\qquad+\frac{A^2N {\rm sign}(h)-b^2 h}{A^2 N^2} \bigg], \label{eq_H}\\
b'' & = & \bigg[\frac{\alpha^2}{A^2 N^2} \hat{r}b'\left(A^2 N^2 h'^2 + b^2 h^2\right)\nonumber\\ & &\qquad-\frac{2 b'}{\hat{r}} + \frac{b h^2}{N}\bigg] . \label{eq_b}
\end{eqnarray}

For the metric function $A(\hat{r})$ we choose the 
boundary condition
$A(\hat{r}_o)=1 \label{Aro} $,
where $\hat{r}_o$ is the outer radius of the star. 
For constructing globally regular ball-like boson star solutions, we choose:
\begin{eqnarray}
& N(0)=1 \ ,\ \   b'(0)=0 \ ,\nonumber\\ &h'(0)=0\ ,\ \ h(\hat{r}_o)=0\ , \ \  h'(\hat{r}_o)=0 . \label{bcstar}
\end{eqnarray}
In the exterior region $\hat{r}>\hat{r}_o$ we match
the Reissner-Nordstr\"om solution. 

The theory has a conserved Noether current:
\begin{eqnarray}
j^\mu=-i \,e\,\left[ \Phi(D^\mu \Phi)^*-\Phi^* (D^\mu \Phi) \right]\ ,\   \
j^{\mu}_{\ ;\mu} = 0\ .
\end{eqnarray}
The charge $Q$ of the boson star is given by
\begin{equation}
\hspace{-0.8cm}Q=-\frac{1}{4\pi}\int_{0} ^{\hat{r}_o} j^t \sqrt{-g} \,dr\,d\theta\,d\phi  \,,\  
j^t=-\frac{h^2(\hat{r}) b(\hat{r})}{A^2(\hat{r}) N(\hat{r})} . \nonumber
\end{equation}
For all boson star solutions we obtain the mass $M$ 
(in the units employed):
 \begin{equation}
M= \biggl(1-N(\hat{r}_o)+\frac{\alpha^{2} Q^{2}}{\hat{r}_o^2}\biggr)\frac{\hat{r}_o}{2} .
 \end{equation}

\begin{figure}
\begin{center}
	\mbox{\subfigure[][]{\includegraphics[width=0.86\linewidth,height=0.17\textheight]{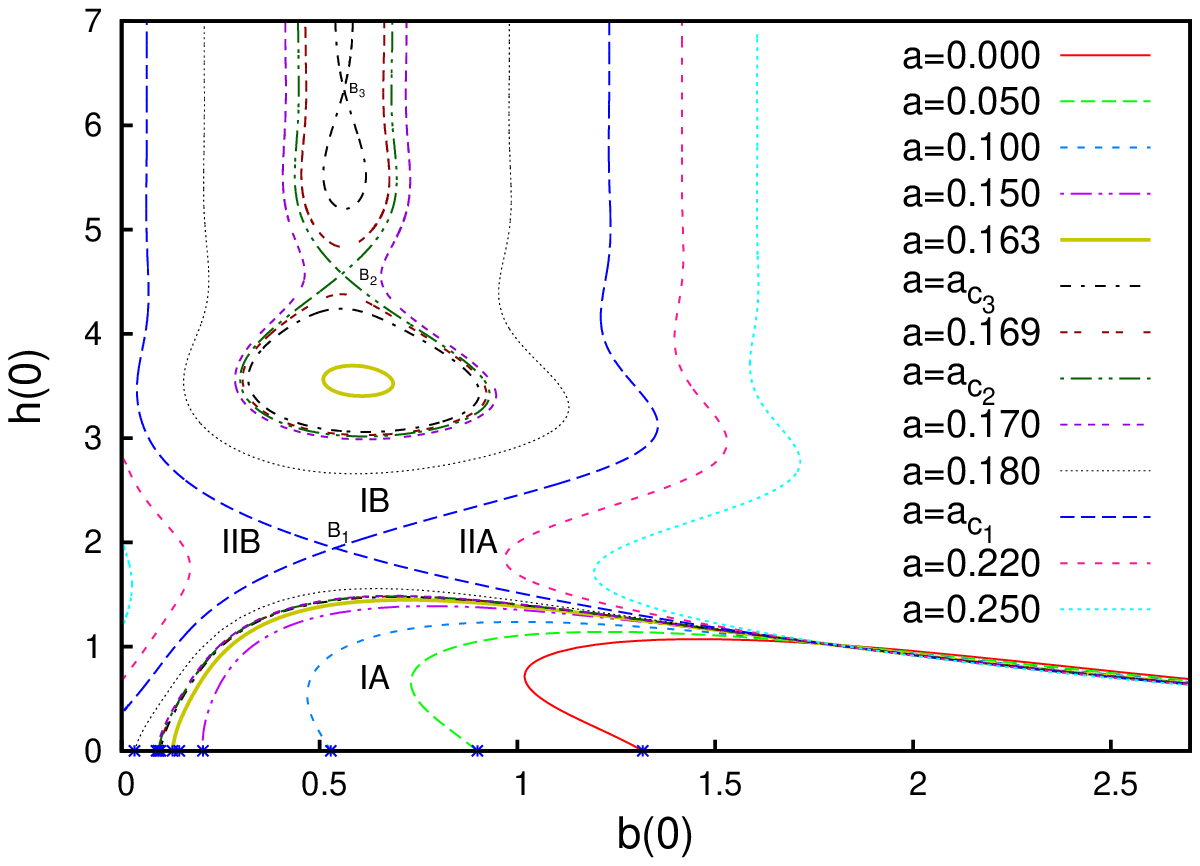}\label{f1a}}}
        \mbox{\subfigure[][]{\includegraphics[width=0.9\linewidth,height=0.17\textheight]{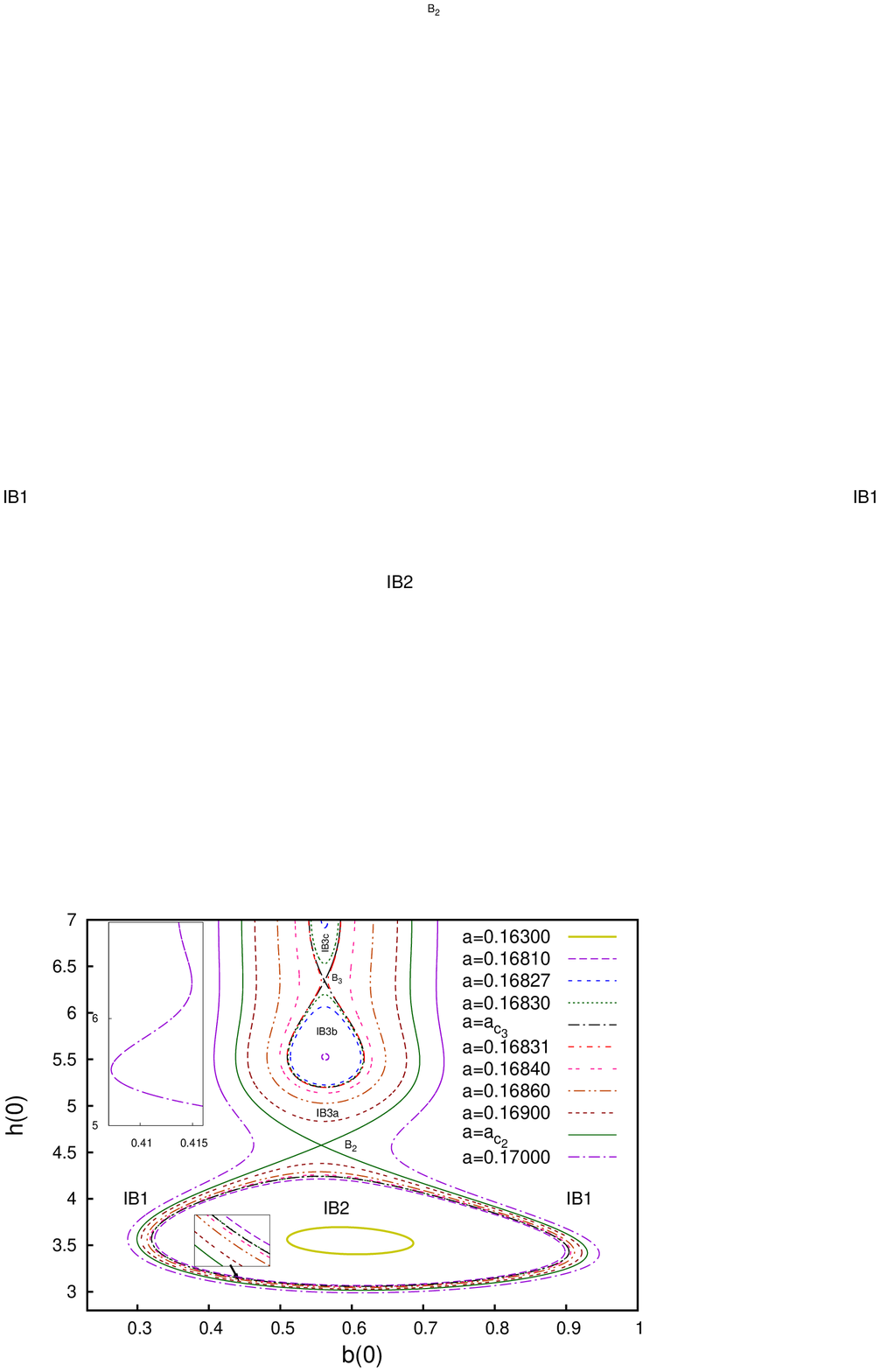}\label{f1b}}}
        \mbox{\subfigure[][]{\includegraphics[width=0.9\linewidth,height=0.169\textheight]{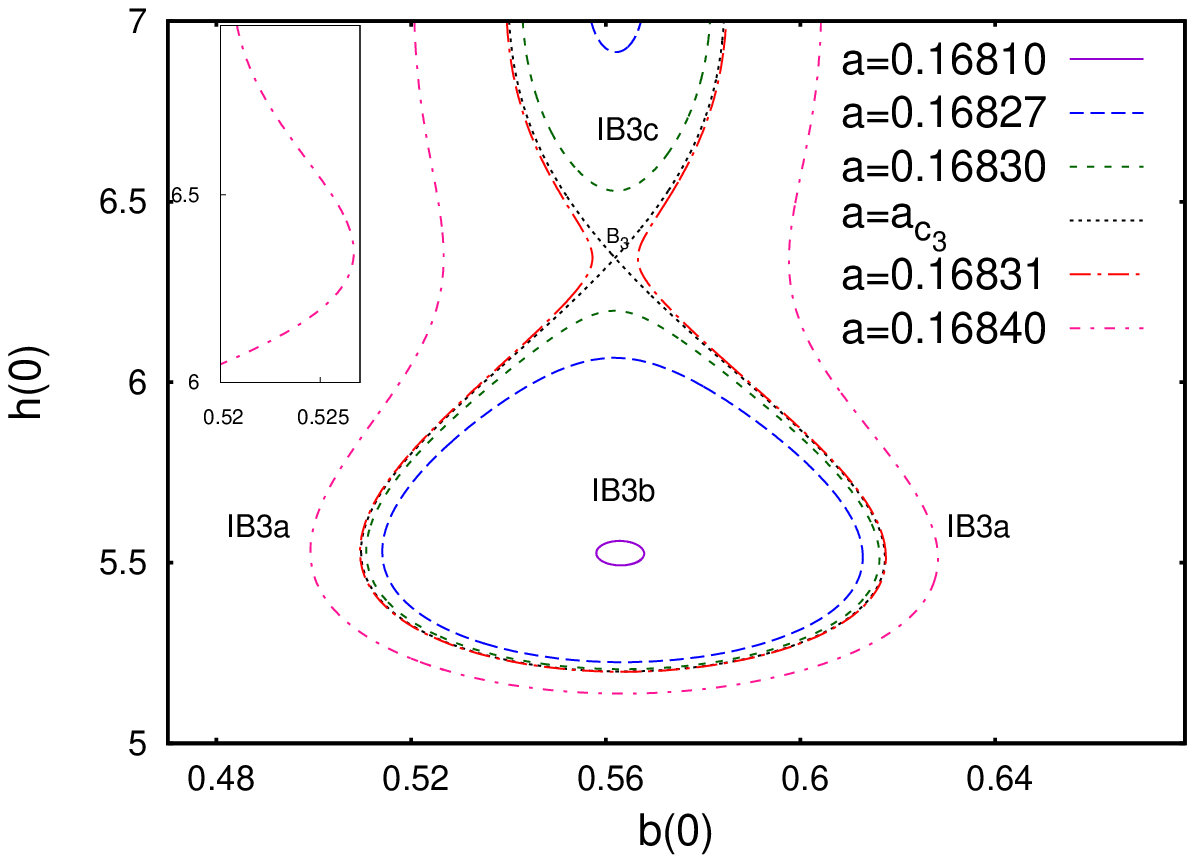}\label{f1c}}}
 	\caption{ Fig.~(a) depicts the phase diagram of the theory 
in terms of the vector field  at the center of the star $b(0)$ 
and the scalar field at the center of the star $h(0)$ 
for different values of the parameter $a$ in the range $a=0$ to $a=0.225$. 
The points $B_1,~B_2 $ and $B_3$ represent three bifurcation points. 
The entire region depicted in the phase diagram in Fig.~(a) 
is divided into four regions IA, IB and IIA, IIB in the vicinity of $B_1$. 
The region IB of the phase diagram shown in Fig.~(a) 
is separately depicted in detail in Fig.~(b). 
The region IB of the phase diagram is subdivided into three regions 
IB1, IB2 and IB3 in the vicinity of $B_2$. 
The region IB3 of the phase diagram shown in Fig.~(b) 
is separately depicted in detail in Fig.~(c). 
It is subdivided into three regions 
IB3a, IB3b and IB3c in the vicinity of $B_3$. 
The asterisks shown in Fig.~(a), corresponding to $h(0)=0$, 
represent the transition points from the boson stars to boson shells. 
The insets in Figs.~(b) and (c) represent 
parts of these phase diagrams with higher resolution.
\label{fig1}}
\end{center}
\end{figure}
\begin{figure}
\begin{center}
	\mbox{\subfigure[][]{\includegraphics[width=0.9\linewidth,height=0.19\textheight]{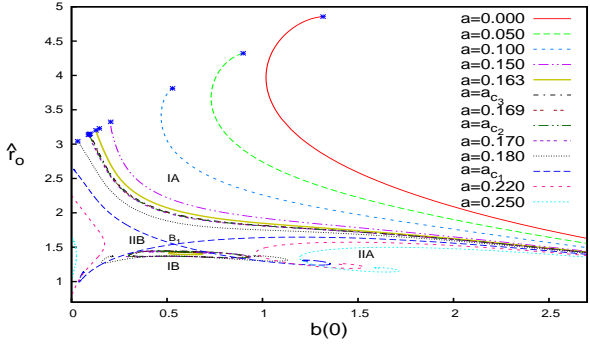}\label{f2a}}}
	\mbox{\subfigure[][]{\includegraphics[width=0.9\linewidth,height=0.19\textheight]{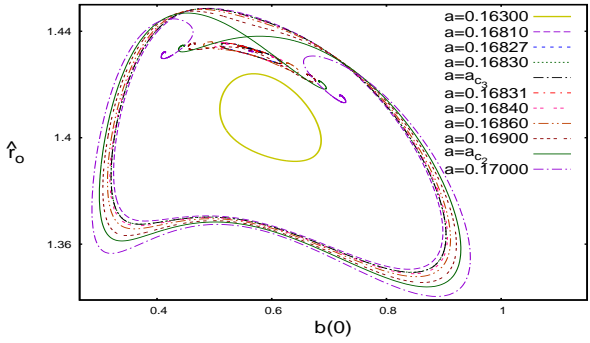}\label{f2b}}}
	\mbox{\subfigure[][]{\includegraphics[width=0.9\linewidth,height=0.19\textheight]{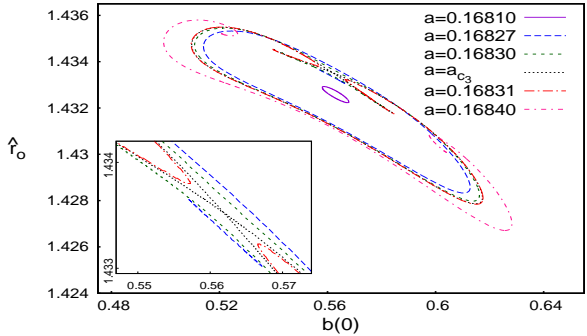}\label{f2c}}}
	\caption{Fig.~(a) shows the radius $\hat{r}_o$
of the boson star versus the vector field 
at the center of the star $b(0)$ for different values of $a$ 
in the range $a=0$ to $a=0.225$. 
The point $B_1$ corresponds to the first bifurcation point 
and the entire region depicted in Fig.~(a) 
is divided into four regions IA, IB and IIA, IIB 
in the vicinity of $B_1$ (as in Fig. \ref{fig1}). 
The region IB shown in Fig.~(a) is separately depicted 
in detail in Fig.~(b) and similarly a part of the region 
shown in Fig.~(b) is separately depicted in detail in Fig.~(c). 
The asterisks shown in Fig.~(a) represent the transition points 
from the boson stars to boson shells. 
The spiral behaviour of the solutions 
is visible in the regions IA and IIB. 
The inset in Fig.~(c) represents a part of the region IIB 
with higher resolution. \label{fig2}}
\end{center}
\end{figure}

We now study the numerical solutions of Eqs.~(\ref{eq_N})-(\ref{eq_b}) 
with the boundary conditions defined by 
$A(\hat{r}_o)=1$ and Eq.~(\ref{bcstar}),
and determine their domain of existence for a sequence
of specific values of the parameter $a$.

Let us recall here that the theory defined by the action 
(Eq. (\ref{3:action})) originally has two parameters $e$ and $\lambda$ 
which are the two coupling constants of the theory. 
At a later stage we have introduced the
new parameters $\beta$ and $a(:=\alpha^2)$,
and we have rescaled the radial coordinate and the matter functions.
Then the parameter $\beta$ does not appear in the resulting set of
equations (\ref{eq_N})-(\ref{eq_b}). 
Thus the numerical solutions of these coupled differential equations 
can be studied by varying only one parameter, namely $a$.

We first consider the phase diagram of the theory 
based on the values of the fields at the origin of the boson star,
the vector field, $b(0)$, and the scalar field, $h(0)$, 
obtained by studying a sequence of values of the parameter $a$. 
We observe very interesting phenomena near specific values of $a$,
where the system is seen to have bifurcation points 
$B_1\,,\,B_2$ and $B_3$. 
These correspond to the following values of $a$: 
$a_{c_1}\simeq0.198926, ~a_{c_2}\simeq0.169311$ 
and $a_{c_3}\simeq0.168308 $, respectively, 
and the possibility of further bifurcation points is not ruled out. 
Thus the theory is seen to possess rich physics 
in the domain $a = 0.22$ to $a \simeq + 0.16\,$.

For a clear discussion, we divide the phase diagram 
in the vicinity of $B_1$
into four regions denoted by IA, IB, IIA and IIB 
(as seen in Fig.~\ref{f1a}). 
The asterisks seen in Fig.~\ref{f1a}, coinciding with the axis $b(0)$ 
(which corresponds to $h(0)=0$), 
represent the transition points from the boson stars to boson shells
\cite{Kleihaus:2009kr,Kleihaus:2010ep}.

The regions IA, IIA and IIB do not have any further bifurcation points. 
However, the region IB is seen to contain rich physics 
as evidenced by the occurrence of more bifurcation points in this region. 
For better detail, the region IB is magnified in Fig.~\ref{f1b}. 
The region IB is then further divided into 
the regions IB1, IB2 and IB3 in the vicinity of $B_2$,
as seen in Fig.~\ref{f1b}. 

The region IB3 finally is seen to have the further bifurcation point $B_3$. 
In the vicinity of $B_3$ we therefore further subdivide 
the phase diagram into the regions IB3a, IB3b and IB3c,
as seen in Fig.~\ref{f1c}. 
The region IB3b is seen to have closed loops 
and the behaviour of the phase diagram in this region 
is akin to the one of the region IB2. 
Also, the insets shown
in Figs.~\ref{f1b} and \ref{f1c} represent parts 
of the phase diagram with higher resolution.

The figures demonstrate, that
as we change the value of $a$ from $a=0.225 $ to $a=0$, 
we observe a lot of new rich physics. 
While going from $a=0.225 $ to the critical value $a=a_{c_1}$, 
we observe that the solutions exist in two separate domains,
IIA and IIB (as seen in Fig.~\ref{f1a}). 
However, as we decrease $a$ below $a =a_{c_1}$, 
the solutions of the theory are seen to exist in the regions IA and IB 
(instead of the regions IIA and IIB). 
For the sake of completeness it is important to emphasize here,
that the physics in the domain corresponding 
to the values of $a$ larger than $a=0.225$ 
conceptually remains the same as described by the value $a=0.225\,$.

As we decrease the value of $a$ 
from the first critical value $a=a_{c_1}$ 
to the next critical value $a=a_{c_2}$, 
we notice that the region IA in the phase diagram 
shows a continuous deformation of the curves,
and the region IB is seen to have its own rich physics 
as explained in the foregoing. 

As we decrease $a$ below $a_{c_2}$, 
we observe that in the region IA there is again 
a continuous deformation of the curves all the way down to $a=0$. 
However in the region IB, we encounter another bifurcation point,
which divides the region IB into IB1, IB2 and IB3. 
We observe that in the region IB1 there is a continuous deformation 
of the curves, and the region IB2 contains closed loops of the curves. 
The region IB3 is subdivided into the regions IB3a, IB3b and IB3c. 
The region IB3a would have a continuous deformation 
of the curves, and the region IB3b is seen to contain closed loops. 

It is tempting to conjecture, that there is a whole
sequence of further bifurcation points, 
leading to a self-similar pattern of the new subregions involved.
The numerical calculations, however, become more and more
challenging, as one proceeds from the first
to the higher bifurcations, since an increasing numerical accuracy
is necessary to map out the domain of existence.
Note, that the value of $a$ had to be specified to
6 decimal digits for B2 and B3, already.
Thus it is the global accuracy of the scheme, 
which presents a limiting factor. Within this accuracy,
the Newton-Raphson method will provide a new solution,
when an adequate starting solution 
has been specified, though.

A plot of the radius $\hat{r}_o$ of the solutions versus 
the vector field at the center of the star $b(0)$ 
is depicted in Fig.~\ref{f2a}. 
As before, the point $B_1$ corresponds to the first bifurcation point,
and the four regions IA, IB and IIA, IIB in the vicinity 
of the bifurcation point are indicated. 
Again, the region IB shown in Fig.~\ref{f2a} 
is enlarged and shown in Fig.~\ref{f2b},
with the region IB3 being enlarged further and
depicted in Fig.~\ref{f2c}.
The asterisks shown in Fig.~\ref{f2a} 
again represent the transition points from the boson stars to boson shells. 
The oscillating behaviour seen in Figs.~\ref{f1a} to \ref{f1c}
in the regions IIA and IB
translates in the Figs.~\ref{f2a} to \ref{f2c}
into a spiral behavior.
The inset in Fig.~\ref{f2c} represents a part of the region IB 
with higher resolution.

\begin{figure}
\begin{center}
	\mbox{\subfigure[][]{\includegraphics[width=0.88\linewidth,height=0.22\textheight]{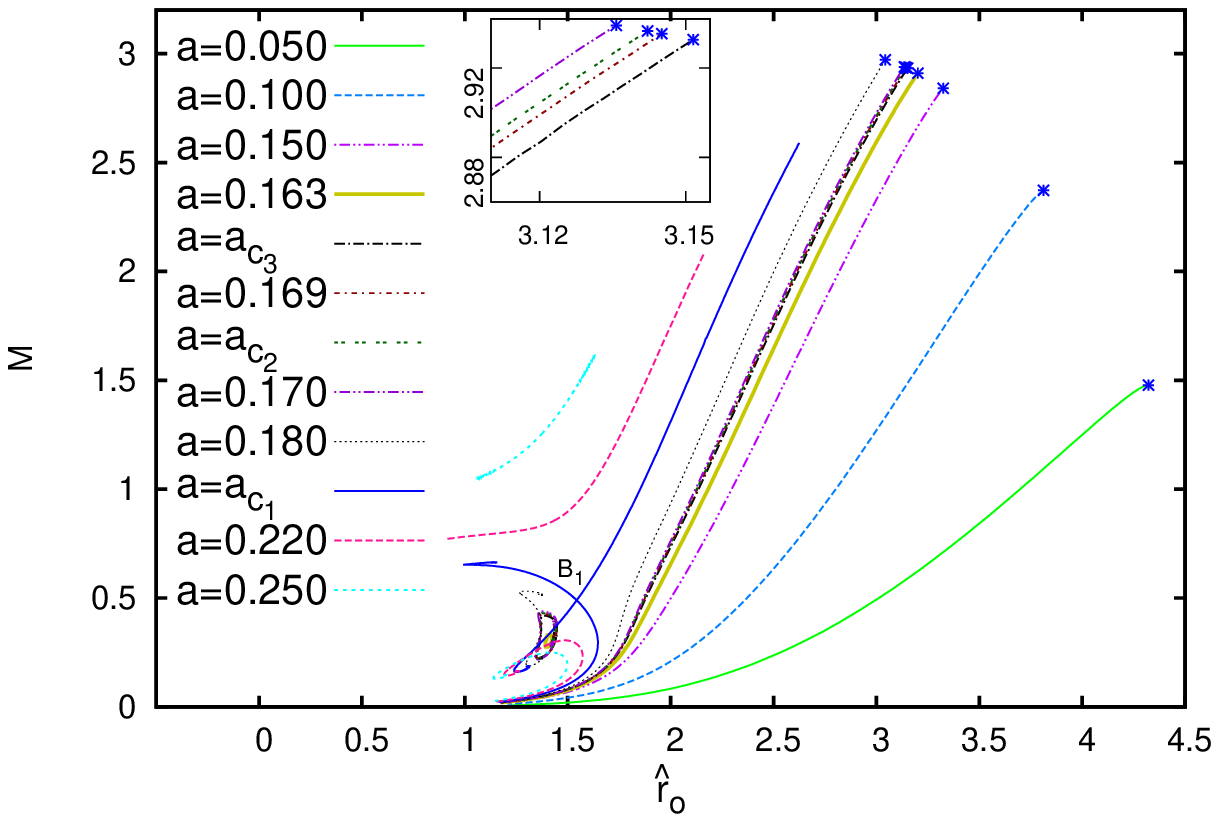}\label{f3a}}}
        \mbox{\subfigure[][]{\includegraphics[width=0.9\linewidth,height=0.22\textheight]{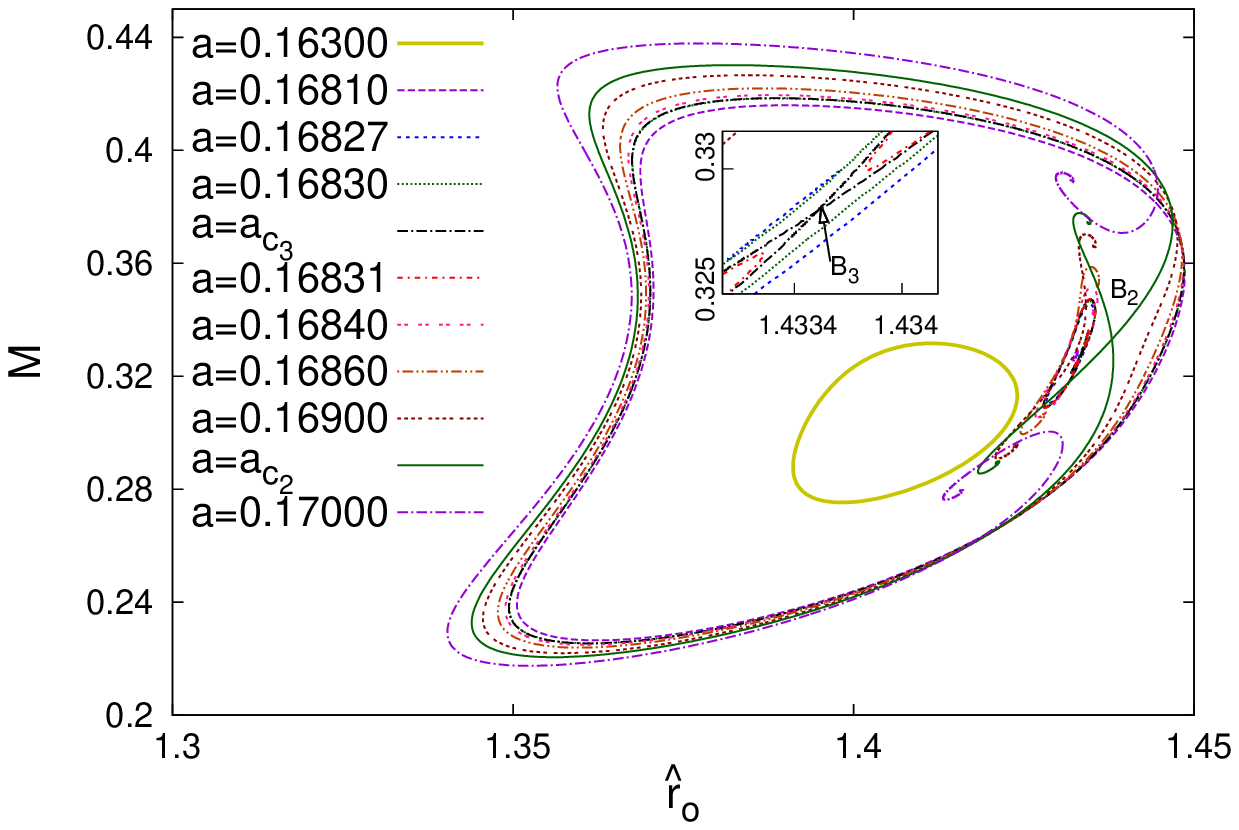}\label{f3b}}}
        \mbox{\subfigure[][]{\includegraphics[width=0.9\linewidth,height=0.22\textheight]{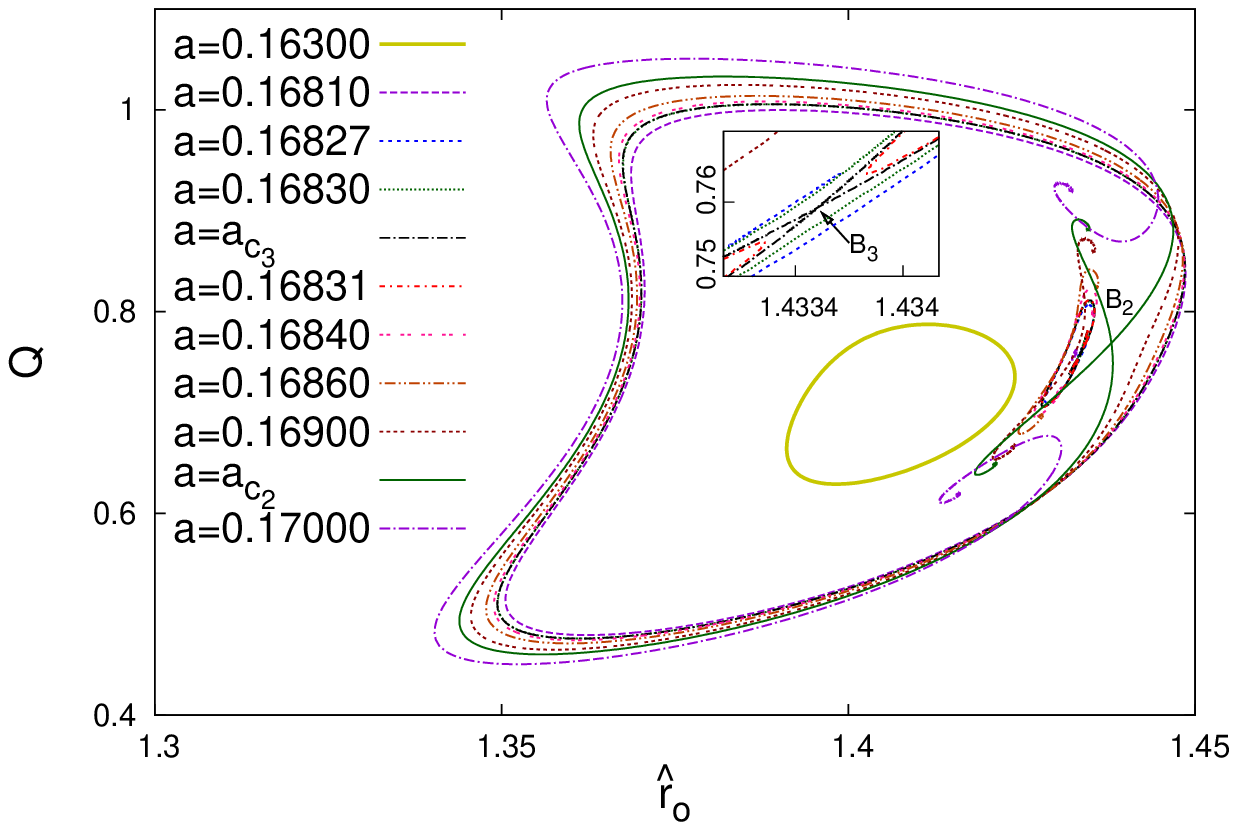}\label{f3c}}}
 	\caption{Fig.~(a) depicts the mass $M$ 
versus the radius of the star $\hat{r}_{o} $ for the
same sequence of values of the parameter $a$.
As before, the asterisks represent the transition points from 
the boson stars to boson shells, and the insets
magnify parts of the diagram.
Fig.~(b) zooms into the region of the bifurcations,
with the inset giving a magnified view of the bifurcation B3.
Fig.~(c) is the analog of Fig.~(b) for the charge $Q$.
\label{fig3}}
\end{center}
\end{figure}

Let us now turn to the global properties of the solutions,
their mass $M$ and their charge $Q$.
The mass $M$ versus the radius $\hat{r}_{o}$ 
is shown in Fig.~\ref{f3a}, while \ref{f3b} again magnifies the
region of the bifurcations.
The charge $Q$ has a very similar dependence as the mass.
This is illustrated in Fig.~\ref{f3c} for the bifurcation region.

\begin{figure}
\begin{center}
\mbox{\subfigure[][]{\includegraphics[width=0.88\linewidth,height=0.22\textheight]{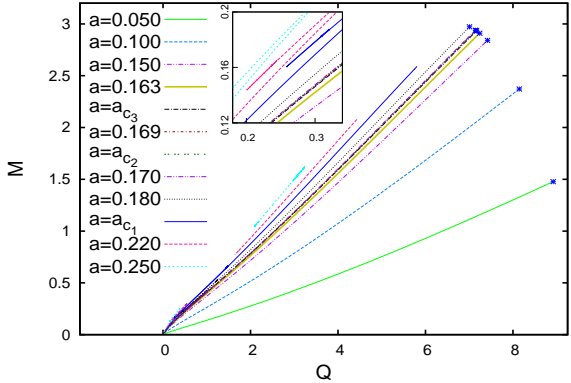}\label{f4a}}}
        \mbox{\subfigure[][]{\includegraphics[width=0.9\linewidth,height=0.22\textheight]{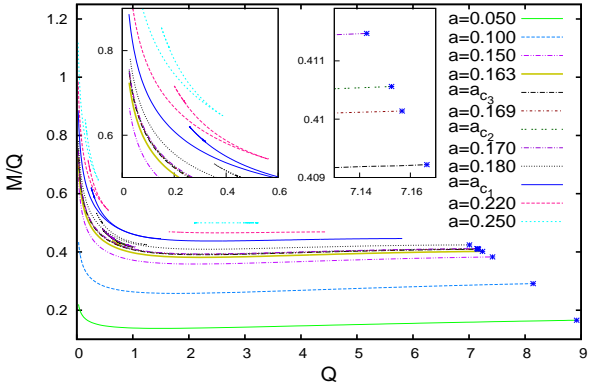}\label{f4b}}}
        \mbox{\subfigure[][]{\includegraphics[width=0.9\linewidth,height=0.22\textheight]{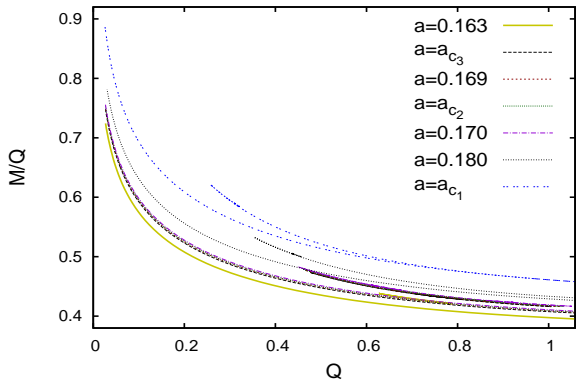}\label{f4c}}}
 	\caption{ Fig.~(a) depicts the mass $M$ versus the charge $Q$
for the same set of solutions. 
As before, the asterisks represent the transition points from
the boson stars to boson shells, and the inset
magnifies a part of the diagram.
Fig.~(b) depicts the mass per unit charge $M/Q$ versus the charge $Q$.
Again the insets magnify parts of the diagram.
Fig.~(c) zooms further into the region of the bifurcations.
\label{fig4}}
\end{center}
\end{figure}

To understand the stability of the boson stars, one can consider
the mass $M$ versus the charge $Q$, as shown in Fig.~\ref{f4a},
or the mass per unit charge $M/Q$ versus the charge,
as shown in Figs.~\ref{f4b} and \ref{f4c}.
Let us first consider Fig.~\ref{f4a}.
Here the curves $M$ versus $Q$, corresponding to the region IA 
and the smaller values of $a$, all increase monotonically from 
$M=Q=0$ to the respective transition points with boson shells,
marked by the crosses. The solutions on these curves can be
considered as the fundamental solutions for their respective
value of $a$. Thus they should be stable.
In fact, all curves in region IA should be stable,
representing the solutions with the lowest mass for a given charge
(and parameter $a$). However, above a certain
value of $a$, these curves no longer reach a boson shell,
but instead their upper endpoint represents a
solution, where a throat is formed. 
The exterior space-time $r>r_0$ then corresponds 
to the exterior of an extremal RN space-time.
This happens whenever the value $b(0)=0$
is encountered, as discussed in detail
previously \cite{Kleihaus:2009kr,Kleihaus:2010ep}.

For the curves shown in region IIB both endpoints correspond
to solutions with throats, since at both endpoints
$b(0)=0$ is encountered. Since these solutions
also represent the lowest mass solutions for a given charge,
they should be stable as well.
In the region IIA, however, the solutions exhibit the
typical oscillating/spiral behavior known for non-compact
boson stars.
In a mass versus charge diagram, this translates into the
presence of a sequence of spikes, as seen in the insets
of Figs.~\ref{f4a} and \ref{f4b}.
Here the solutions should be stable only on their fundamental
branch, reaching up to a maximal value of the mass and the charge,
where a first spike is encountered. With every following spike
a new unstable mode is expected to arise, as we conclude
by analogy with the properties of non-compact boson stars.

In this work our focus has been on the bifurcations.
Let us therefore now inspect the region of the bifurcations IB,
starting with the limiting curves.
For the value $a_{c_1}$ 
the two branches of solutions, limiting the region IA,
possess lower masses than the 
the two branches of solutions, limiting the region IB,
and should therefore be more stable.
The two branches of solutions, limiting the region IB,
might be classically stable as well, until the first
extrema of mass and charge are encountered.
Quantum mechanically, however, they would be unstable,
since tunnelling might occur.
Beyond these extrema, unstable modes should be present,
and thus the solutions should also be classically unstable.

These arguments can be extended to all the solutions
in region IB. From a quantum point of view they should be
unstable, since for all of them there exist solutions
in region IA,
which have lower masses but possess the same values of the charge.
Classically, however, the lowest mass
solutions for a given $a$ within the region IB might
be stable, while the higher mass solutions should
clearly possess unstable modes and be classically unstable.
Fig.~\ref{f4c} zooms into the bifurcation region of
the $M/Q$ versus $Q$ diagram, to illustrate
that the solutions in the bifurcation region 
indeed correspond to higher mass solutions.

In conclusion, we have studied in this work 
a theory of a complex scalar field with a conical potential,
coupled to a U(1) gauge field and gravity 
\cite{Kleihaus:2009kr,Kleihaus:2010ep}. 
We have constructed the boson star solutions of this theory 
numerically and investigated their domain of existence,
their phase diagram, and their physical properties.

We have shown that the theory has rich physics 
in the domain $a=0.22$ to $a\simeq0.16$, 
where we have identified three bifurcation points $B_1,~B_2$ and $B_3$
of possibly a whole sequence of further bifurcations.
We have investigated the physical properties of the solutions,
including their mass, charge and radius.
By considering the mass versus the charge (or the mass per unit
charge versus the charge) we have given arguments concerning
the stability of the solutions.

For all values of $a$ studied, there is a fundamental branch
of compact boson star solutions, which should be stable,
since they represent the solutions with the lowest mass
for a given value of the charge,
and thus represent the ground state.
In the region of the bifurcations additional branches
of solutions are present, which possess higher masses
for a given charge. 
Thus these solutions correspond to excited states
of the system.
The lowest of these might be classically stable, as well,
and only quantum mechanically unstable.
To definitely answer this question, a mode stability analysis
should be performed, which is, however, beyond the scope
of this paper, representing a topic
of separate full-fledged investigations.

Finally, we would like to mention that detailed investigations 
of this theory in the presence of the cosmological constant $\Lambda$ 
with 3D plots of the phase diagrams involving 
the various physical quantities of the theory 
are currently under our investigation 
and would be reported later separately.


We would like to thank James Vary for very useful discussions. 
This work was supported in part by the US Department of Energy 
under Grant No. DE-FG02-87ER40371,
by the US National Science Foundation under Grant No. PHY-0904782,
by the DFG Research Training Group 1620
{\sl Models of Gravity} as well as by FP7, Marie Curie Actions, People
IRSES-606096.
SK would like to thank the CSIR, New Delhi, 
for the award of a Research Associateship.

\end{document}